\documentclass[doublecol]{epl2}
\usepackage{graphicx}

\title{Active Rheology in Odd Viscosity Systems}
 
\author{ C. J. O. Reichhardt and C. Reichhardt}
\institute{
Theoretical Division and Center for Nonlinear Studies,
Los Alamos National Laboratory, Los Alamos, New Mexico 87545, USA}

\abstract{
Odd viscosity arises in systems with time reversal symmetry breaking, which creates non-dissipative effects. One method to probe changes in viscosity is to examine the dynamics of a single probe particle driven though a medium, a technique known as active rheology. We show that active rheology in a system with odd viscosity and no quenched disorder reveals a variety of novel effects, including a speed up of the probe particle with increasing system density when the background medium creates a velocity boost of the driven particle due to the Magnus effect. In contrast, the probe particle velocity in the dissipation-dominated limit monotonically decreases with increasing system density. We also show that the odd viscosity imparts a Hall angle to the probe particle, and that both the Hall angle and the velocity boost depend strongly on the drive. These results should be general to other systems with odd viscosity, including skyrmions in chiral magnets.        
}

\begin{document}

\maketitle

\section{Introduction}
Viscosity in fluids or glasses is associated with 
dissipation,
and
an increase in density generally produces an increase of viscosity.
For example, when a glass transition is approached from below by
increasing the density, the viscosity increases
\cite{Mauro09}. 
An effective method for examining changes in the viscosity of a medium
is active rheology, where the drag or fluctuations 
on a single probe particle
are measured as the particle is pushed through the
medium
while the density or driving force
is varied \cite{Habdas04,Squires05,Wilson11a,Voigtmann13,Zia18}.
Active rheology has been used to study viscosity changes
in systems undergoing glass
\cite{Habdas04,Wilson11a,Voigtmann13,Gazuz09,Winter12,Yu20} or 
jamming \cite{Drocco05,Candelier10,Kolb13} transitions.  
The probe particle velocity
decreases or even
drops to zero
at a pinning transition when the driving force drops below a 
critical value or the
density rises above a critical value
\cite{Habdas04,Voigtmann13,Drocco05,Gruber16,Senbil19,Gruber20}.
The corresponding viscosity increase
is produced by
the increased interactions with the
background particles,
since the probe particle must
push more particles out of its way in order to 
translate in the driving direction.
Active rheology has also been used to study viscous properties
in active matter systems \cite{Reichhardt15}, 
depletion interactions \cite{Wulfert17},
local melting \cite{Dullens11},  
and complex soft matter systems \cite{Wang19,AbaurreaVelasco20} 

In systems where time reversal symmetry breaking occurs,
another type of viscosity 
termed odd viscosity 
can arise.
Here, the Onsager reciprocal relations break down,
there are no energy eigenstates,
and nondissipative effects can arise
due to the leaking of energy from one mode to another.
Odd viscosity is also called gyroviscosity since it can appear in systems 
with gyroscopic effects such as charged particles in magnetic fields, 
quantum Hall fluids \cite{Avron95,Avron98},
plasmas \cite{Lifshitz81},
fluid vortices \cite{Wiegmann14},
and non-dissipative systems with broken parity \cite{Abanov20}.
More recently, odd viscosity has been
studied in chiral active matter systems
or active spinners \cite{Banerjee17,Soni19,Souslov19,Hargus20,Hosaka21} 
where it
generates flows that are perpendicular to pressure \cite{Banerjee17},
unidirectional propagation of edge modes \cite{Soni19},
topological waves \cite{Souslov19}, and anomalous flows
\cite{Markovich20}.
Recently, similar effects have also been explored
in systems that have odd elasticity \cite{Scheibner20} or exhibit 
non-reciprocal phase transitions \cite{Reichhardt21,Fruchart21}. 

Here we show that odd viscosity can have pronounced
effects on active rheology,  
producing significantly different and even
reversed behavior compared
to that found
in systems with only dissipative viscosity.  One of the most prominent 
distinctions
is that the velocity of the probe particle
at a constant drive can {\it increase} with increasing
system density,
which is the opposite of
what is observed in systems with dissipative viscosity. 
Another consequence is that
the probe particle
exhibits a finite Hall angle with a value that depends
on the driving force
and the density.
A speed up effect appears as a function of increasing density
when a density gradient is built up
by the finite Hall angle motion of the probe particle.   
The odd viscosity
transfers the pressure from the
unbalanced density gradient forces
into a flow perpendicular to these forces and parallel to the
driving direction.
The speed up effect
has a non-monotonic dependence on the driving force
and produces
velocity-force signatures distinct from those
found in the overdamped limit. 
The specific model we consider represents
skyrmions in chiral magnets \cite{Muhlbauer09,Yu10,Nagaosa13}
which have a strong Magnus force component 
due to their topology \cite{Nagaosa13,EverschorSitte14,Reichhardt21a}.
Our results indicate that skyrmions are another system that
exhibits strong odd-viscosity effects,
and our findings should be general to
the broader class of systems with odd viscosity.   

\begin{figure}
  \onefigure[width=0.75\columnwidth]{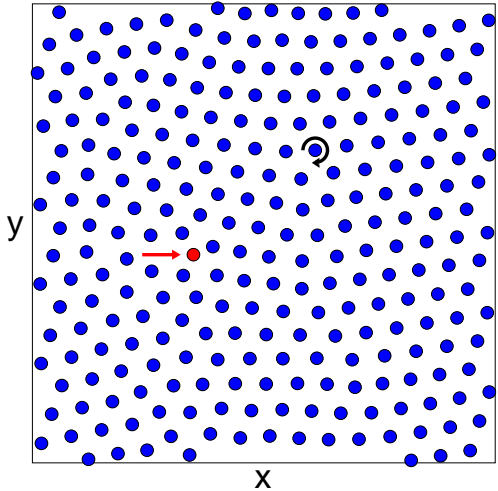}
\caption{(a) Image of a portion of the sample at
density $\rho = 0.2$. The red particle is
being driven in the $x$-direction (red arrow) at a constant force $F_{D}$,
and it interacts with the surrounding bath particles (blue).
The black arrow around a representative bath particle indicates
the direction of chirality 
of the Magnus term. 
}
\label{fig:1}
\end{figure}

\section{Simulation and System}
In Fig.~\ref{fig:1} we show a snapshot of
a portion of our two-dimensional
system, which has
periodic boundary conditions in the $x$ and $y$ directions.  
A probe particle is subjected
to a constant driving force of
${\bf F}_{D}=F_D{\bf \hat{x}}$
and
interacts with a bath of other particles which
experience both dissipative and Magnus forces. The 
chirality of the Magnus force is
indicated by the circulating arrow around a representative bath particle.
The dynamics of particle $i$ obey
the following equation of motion:
\begin{equation}  
\alpha_d {\bf v}_{i} + \alpha_m {{\bf \hat z}} \times {\bf v}_{i} = 
{\bf F}^{ss}_{i} + {\bf F}^{D} 
\end{equation}
Here ${\bf v}_i$ (${\bf r}_i$) is the velocity (position) of particle $i$.
The dissipative damping constant $\alpha_d$
aligns the velocity in the direction of the net applied forces, 
while the Magnus term $\alpha_m$ creates a velocity
component perpendicular to the net forces
and produces
the odd-viscosity behavior.
The Magnus term can arise from 
spinning of the particles, cyclotron motion,
or the intrinsic topological nature of the particle
as in the case of magnetic skyrmions \cite{Nagaosa13,EverschorSitte14}. 
The  particle-particle interaction force is 
${\bf F}^{ss}_{i} = \sum^{N}_{j=1} K_{1}(r_{ij}) \hat{\bf r}_{ij}$
where $r_{ij}=|{\bf r}_i - {\bf r}_j|$,
$\hat{\bf r}_{ij}=({\bf r}_i - {\bf r}_j)/r_{ij}$, and $K_{1}$
is the modified Bessel function
which gives a smoothly decreasing force
of the form $e^{r}/r$ to create an intermediate range repulsive interaction. 
This interaction has been used previously
to model skyrmions in chiral magnets \cite{Lin13};
however, the smooth repulsive interaction should also be relevant
for other
systems with spinning objects or some form of Magnus force,
including spinning magnetic or charged 
colloids, fluid vortices, fractons, Wigner crystals in a magnetic field,
or certain types of chiral active matter. 
The driving force ${\bf F}_{D}=F_D{\bf \hat{x}}$ is applied only  
to the probe particle.
We measure the probe particle
velocity parallel,
$\langle V_{||}\rangle$, and perpendicular,
$\langle V_{\perp}\rangle$ to the drive, where the average is taken over time,
as well as
the net velocity
$|V|  = (\langle V_{\perp}\rangle^2 + \langle V_{||}\rangle^2)^{1/2}$.  
In the absence of any collisions, the probe particle
has an
intrinsic Hall angle of
$\theta^{\rm int}_{\rm Hall} = -\arctan(\alpha_{m}/\alpha_{d})$,
while in the presence of the background particles, we measure
the actual Hall angle
$\theta_{\rm Hall} = \arctan(\langle V_{\perp}\rangle/\langle V_{||}\rangle).$
In the damping-dominated limit of $\alpha_m=0$,
$\theta^{\rm int}_{\rm Hall} = 0^{\circ}$.
In skyrmion systems
this is referred to as the
skyrmion Hall angle \cite{Nagaosa13,EverschorSitte14,Reichhardt21a}
and it has been observed directly in experiment
\cite{Jiang17,Litzius17,Reichhardt20,Juge19,Zeissler20}. 

{\it Results---}
We first  consider a constant driving force of $F_{D} = 1.0$ for varied 
system density $\rho$.
In order to more easily compare systems with different ratios of
damping to Magnus forces,
we impose the constraint $\alpha_{m}^2 + \alpha_{d}^2 = 1.0$. 
In the single particle limit,
$|V| = F_{D}/(\alpha_{m}^2 + \alpha_{d}^2)^{1/2}$,
which under our constraint
gives $|V| = F_{D}$, independent of the ratio of the
force terms.
This makes it easy to determine whether the velocity is decreased or
boosted with respect to the single particle limit as we change $\rho$.

\begin{figure}
  \onefigure[width=\columnwidth]{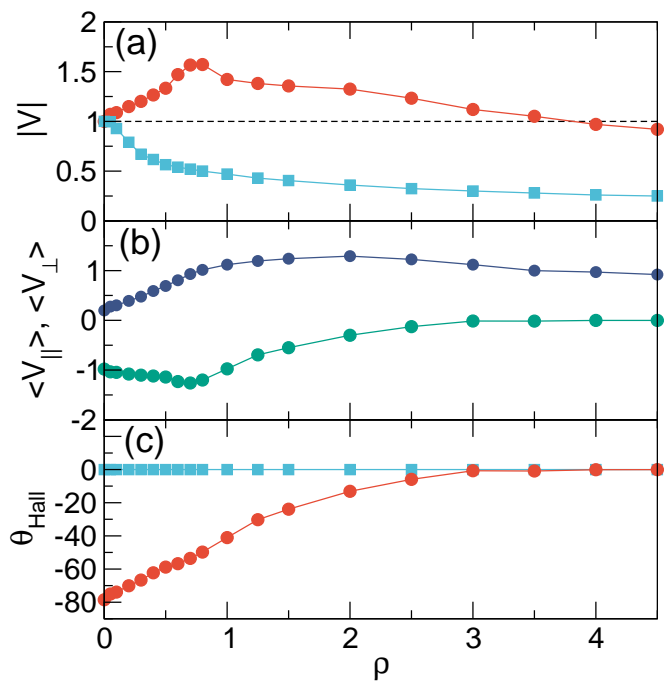}
\caption{(a) The total
probe particle velocity $|V|$ versus density $\rho$ for a system with
$F_{D} = 1.0$
in the overdamped regime with $\alpha_m=0$ (blue squares)
and in the Magnus dominated regime  
with $\alpha_{m}/\alpha_{d} = 4.9246$ (red circles).
The dashed black line is the expected velocity in the single particle limit.
(b) The velocities parallel and perpendicular to the drive,
$\langle V_{||}\rangle$ (blue) and $\langle V_{\perp}\rangle$ (green),
versus $\rho$
for the Magnus dominated system in panel (a).
(c) The corresponding measured skyrmion Hall angle
$\theta_{\rm Hall}$ versus $\rho$ for the overdamped system (blue squares) and
the Magnus dominated system (red circles).}
\label{fig:2}
\end{figure}

In Fig.~\ref{fig:2}(a) we plot
$|V|$ versus $\rho$ for an overdamped system with $\alpha_m=0$
and a Magnus dominated system with $\alpha_{m}/\alpha_{d} = 4.9246$.
In the overdamped limit,
$|V|$ monotonically decreases with increasing $\rho$
since the probe particle must displace a larger number
of particles in order to move through the system.
Here $|V|$ is always lower than the value
$|V|=F_D=1.0$ expected in the single particle limit.
In contrast, for the Magnus dominated system $|V|$ has the
opposite behavior and
increases with increasing density
before reaching a maximum value near $\rho = 0.8$. As $\rho$ increases
further, $|V|$ gradually decreases.
Although $|V|=1$ for low $\rho$,
it rises above this value over an extended range of density.
At $\rho = 0.8$, $|V|$ for the Magnus dominated system is about three times
higher than $|V|$ in the overdamped system.

In Fig.~\ref{fig:2}(b) we plot
$\langle V_{||}\rangle$ and $\langle V_{\perp}\rangle$ for the
$\alpha_{m}/\alpha_{d} = 4.9246$ sample.
In the single particle limit,
$|\langle V_{\perp}\rangle|=4.9246 \langle V_{||}\rangle$.
As the density $\rho$ of the bath particles increases,
$\langle V_{||}\rangle$ increases to a maximum value
near $\rho = 2.0$, 
while the magnitude of $\langle V_{\perp}\rangle$
reaches its largest value near
$\rho = 0.8$. Both quantities decrease as $\rho$ increases further.
For the overdamped case (not shown),
$\langle V_{\perp}\rangle = 0$ and $\langle V_{||}\rangle$ is 
identical to $|V|$.
In Fig.~\ref{fig:2}(c) we plot the Hall angle
$\theta_{\rm Hall}$ versus $\rho$.
In the overdamped limit,
$\theta_{\rm Hall} = 0^\circ$ for all $\rho$.
For the Magnus dominated system, in the single particle limit we would
expect to obtain
$\theta_{\rm Hall} = -\arctan(\alpha_{m}/\alpha_{d}) = -78.52^\circ$,
but when interactions with the bath particles are present,
the magnitude of $\theta_{\rm Hall}$
monotonically decreases with increasing $\rho$ and reaches a value
close to zero for $\rho > 3.0$. 

\begin{figure}
  \onefigure[width=\columnwidth]{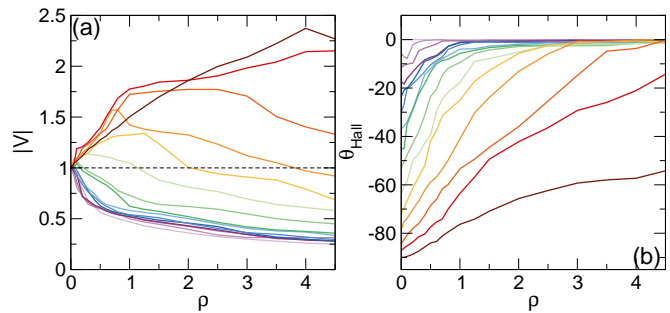}
\caption{(a) $|V|$ versus $\rho$
for the system in Fig.~\ref{fig:2} with $F_D=1.0$ at
$\alpha_{m}/\alpha_{d} = 0.0$, 0.1, 0.204, 0.3145, 0.4364, 0.577, 0.75,
1.0, 1.33, 2.0647, 3.0424, 4.9246, 9.95,
and $19.97$, from bottom to second from top,
where the brown curve is for a system with $\alpha_{d} = 0.0$
and $\alpha_{m} = 1.0$.
The dashed line is the single particle limit of $V_0=F_D=1.0$.
(c) $\theta_{\rm Hall}$ versus $\rho$ for the same system.
}
\label{fig:3}
\end{figure}

In Fig.~\ref{fig:3}(a) we plot $|V|$ versus $\rho$ for the
system in Fig.~\ref{fig:2} with $F_D=1.0$ at 
$\alpha_{m}/\alpha_{d} = 0.0$, 0.1, 0.204, 0.3145, 0.4364, 0.577, 0.75,
1.0, 1.33, 2.0647, 3.0424, 4.9246, 9.95, and $19.97$,
as well as for a system with
$\alpha_{d} = 0.0$ and $\alpha_{m} = 1.0$.
When $\alpha_{m}/\alpha_{d} < 0.75$, $|V|$ decreases
monotonically with increasing $\rho$ and always
falls below
the single particle limit of $V_{0}= 1.0$, indicated by a dashed line.
For $\alpha_{m}/\alpha_{d} > 0.75$, there is a growing window
with $|V|>V_0$
indicating a velocity boost.
At lower $\rho$,
$|V|$ increases with increasing $\rho$;
however, at higher $\rho$, $|V|$ drops below $V_{0}$.
For $\rho=0$ 
all of the curves start at the single particle value of $|V|=V_0$,
indicating that the velocity boosts are
the result of collective interaction effects.
For the systems with the largest
Magnus terms, the window of increasing $|V|$ extends as high
as 
$\rho = 4.0$.
Figure~\ref{fig:3}(b) shows the corresponding
$\theta_{\rm Hall}$ versus $\rho$.
With increasing $\rho$,
the magnitude of $\theta_{\rm Hall}$ monotonically decreases
from its value in the single particle limit,
indicting that the increase in
collision frequency is responsible for the
reduction of the Hall angle.
There is
an extended region where $\theta_{\rm Hall}$
is close to zero at the higher densities. 

\begin{figure}
  \onefigure[width=\columnwidth]{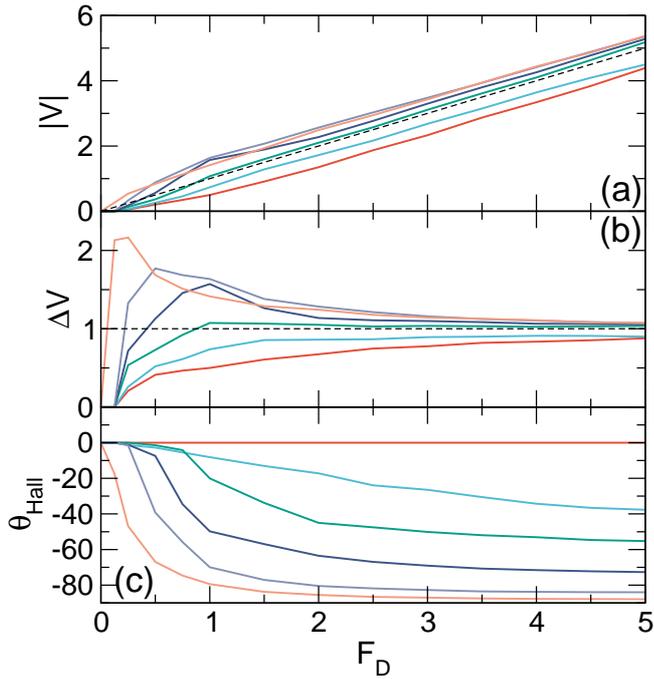}
\caption{(a) $|V|$ versus $F_{D}$ for
systems with $\rho = 0.8$ and 
$\alpha_{m}/\alpha_{d} = 0.0$ (red), $1.0$ (light blue),
$2.0647$ (green), $4.9245$ (dark blue), and $19.97$ (light purple),
with the orange curve indicating a system with
$\alpha_{d}=0$ and $\alpha_{m} = 1.0$.
The dashed line is the
curve for the single particle limit.
(b) The corresponding velocity boost $\Delta V = |V|/V_{0}$
versus $F_{D}$, where $V_{0}$ is the velocity in the single particle limit.
(c) The corresponding $\theta_{\rm Hall}$ versus $F_{D}$. 
}
\label{fig:4}
\end{figure}
 
In Fig.~\ref{fig:4}(a) we plot $|V|$ versus $F_{D}$
for systems with $\rho = 0.8$ at 
$\alpha_{m}/\alpha_{d} = 0.0$, $1.0$, $2.0647$, $4.9245$, and $19.97$,
and for a system with $\alpha_{d} = 0.0$ and $\alpha_{m} = 1.0$.
When $\alpha_{m}/\alpha_{d} < 2.064$,
$|V|$ falls below the single
particle limit,
while for $\alpha_{m}/\alpha_{d} \geq 2.064$,
a portion of the velocity curve rises above the single particle limit,
indicating a boost effect.
In Fig.~\ref{fig:4}(b) we quantify the boost by plotting
$\Delta V = |V|/V_{0}$ versus $F_{D}$,
where $V_{0}$ is the velocity in the single particle limit.
At low drives $\Delta V < 1.0$, indicating increased damping due to
the particle-particle interactions. 
In fact, the probe particle is pinned and unable to move through the
background particles
for $F_{D} < 0.15$.
For $\alpha_{m}/\alpha_{d} < 2.064$,
$\Delta V$ gradually approaches
$\Delta V = 1.0$ with increasing
drive, while for
$\alpha_{m}/\alpha_{d} > 2.064$,
$\Delta V$ passes through a maximum and then decreases
back toward $\Delta V=1.0$ as $F_D$ increases.
For $\alpha_{m}/\alpha_{d} = 19.97$,
the maximum boost is close to $1.75$ times higher than the velocity
in the single particle limit.
In Fig.~\ref{fig:4}(c) we plot the corresponding $\theta_{\rm Hall}$
versus $F_{D}$ showing that
$\theta_{\rm Hall}$ 
monotonically increases with increasing
velocity 
and approaches the single particle limit at large drives.
The drive dependence of the Hall angle  
is similar to what has been
observed for skyrmions moving over quenched disorder,
where the skyrmion Hall angle is zero
near the depinning transition 
and increases with increasing skyrmion velocity
up to its intrinsic value
\cite{Reichhardt21a,Jiang17,Litzius17,Reichhardt20,Juge19,Zeissler20,Reichhardt15a}. 
In our case there is no quenched disorder,
but
scattering occurs during the collisions 
with other particles.
For active rheology in an overdamped system,
the probe particle velocity
is always lower than the single particle limit \cite{Habdas04}. 

The origin of the dependence of the velocity
on $\rho$ and $F_{D}$ is the attempts of the probe particle to
move along
its intrinsic Hall angle
$\theta^{\rm int}_{\rm Hall}  = -\arctan(\alpha_{m}/\alpha_{d})$.
The resulting collisions with the background particles produce
a density build up of $\Delta \rho$ below the probe particle,
perpendicular to the drive direction.
Due to the repulsive
nature of the particle-particle interactions,
the probe particle
experiences an unbalanced force
in
the positive $y$-direction
from the background particles.
Since the Magnus force or the odd
viscosity creates a velocity component perpendicular to the net force,
the probe exhibits enhanced velocity in the positive
$x$ direction, parallel to the drive, of magnitude
$V_{x} \propto  F_{D}\alpha_{d}  + \alpha_{m}\Delta \rho$.
As $\alpha_{m}$ increases, the boost increases.
The local density inhomogeneity
also causes
the probe particle to move at
an angle less than $\theta^{\rm int}_{\rm Hall}$. 
The magnitude of $\Delta \rho$    
depends on the overall particle density.
At large $\rho$, it is difficult to maintain a local disturbance in
the background particle density, reducing $\Delta \rho$ and
reducing the boost,
as shown in Fig.~\ref{fig:3}(a).
At low $F_{D}$, the probe particle is moving slowly
enough that the surrounding particles have time to relax away the
local gradient,
so $\Delta \rho$ is close to zero and the boost is small,
while when the probe particle is moving rapidly at high $F_{D}$,
the surrounding particles do not have time to form a local density gradient
and $\Delta \rho$ is again small, giving
a reduced boost at high velocities, as shown in Fig.~\ref{fig:4}(b).
Velocity boosts
due to the Magnus force have been observed for skyrmions interacting
with a repulsive defect line or edge barrier,
where the force from the barrier is perpendicular to the driving force
but gets converted by the Magnus term into a velocity component in
the driving direction \cite{Iwasaki14,CastellQueralt19}.  
In previous work on a probe particle moving in
an overdamped system where the particles are subjected to chiral ac driving,
the probe exhibited a drive  
dependent Hall angle, but there was no velocity boost \cite{Reichhardt19a}. 

\begin{figure}
  \onefigure[width=\columnwidth]{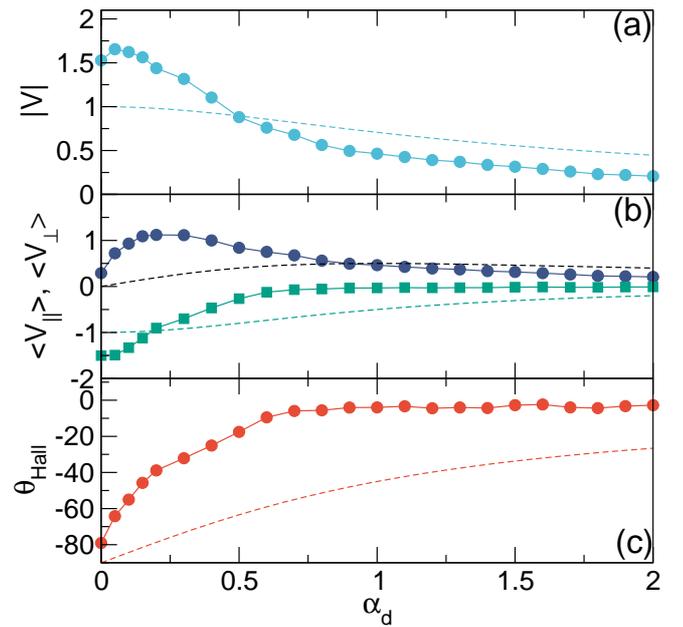}
\caption{(a) $|V|$ versus $\alpha_{d}$ for
a system with $F_{D} = 1.0$ and $\alpha_{m} = 1.0$ at
$\rho = 1.0$ (blue circles)
and the single particle limit of
$|V| = F_{D}/(1 + \alpha^2_{d})^{1/2}$ (dashed line). 
(b) The corresponding
$\langle V_{||}\rangle$ (blue circles)
and $\langle V_{\perp}\rangle$ (green squares).
The single particle limits are 
$\langle V_{||}\rangle = F_{D}\alpha_{d}/(1  + \alpha^2_{d})$
(black dashed line)
and  
$\langle V_{\perp}\rangle = -F_{D}/(1  + \alpha^2_{d})$ (green dashed line).
(c)
$\theta_{\rm Hall} = \arctan(\langle V_{\perp}\rangle/\langle V_{||}\rangle)$
at $\rho = 1.0$ (red circles)  
and the single particle limit of
$\theta_{\rm Hall} = -\arctan(1/\alpha_{d})$ (dashed line).
}
\label{fig:5}
\end{figure}

We next relax the constraint
$\alpha^2_{m} + \alpha^2_{d} = 1.0$
and hold $\alpha_{m}$ fixed while varying $\alpha_{d}$
over the range $0.0$ to $2.0$.
Here, in the single particle limit, 
$|V| = F_{D}/(\alpha^2_{m} + \alpha^2_{d})^{1/2}$,
$\langle V_{||}\rangle = F_{D}\alpha_{d}/(\alpha^2_{m} + \alpha^2_{d})$, and  
$\langle V_{\perp}\rangle = -F_{D}\alpha_{m}/(\alpha^2_{m} + \alpha^2_{d})$.
In Fig.~\ref{fig:5}(a) we plot
$|V|$ versus $\alpha_{d}$
for a system with $\alpha_{m} = 1.0$ and $F_{D} = 1.0$ at
$\rho = 1.0$.
The velocity is boosted
above the single particle limit
for $\alpha_{d} < 0.5$,
while for higher values of $\alpha_{d}$, $|V|$ drops below
the single particle limit, indicating increased damping.
Figure~\ref{fig:5}(b) shows
$\langle V_{||}\rangle$ and $\langle V_{\perp}\rangle$
versus $\alpha_d$
for the $\rho = 1.0$
system.
The magnitudes of the velocities in both directions are first boosted
above the single particle limit, and then drop
below the single particle limit at larger $\alpha_{d}$. 
In Fig.~\ref{fig:5}(c) we plot the corresponding
$\theta_{\rm Hall} = \arctan(\langle V_{\perp}\rangle/\langle V_{||}\rangle)$
versus $\alpha_d$
for the system in Fig.~\ref{fig:5}(b)
along with the single particle limit of
$\theta_{\rm Hall} = -\arctan(\alpha_{m}/\alpha_{d}) = -\arctan(1/\alpha_{d})$.
The magnitude of the Hall angle is reduced
in the interacting system, and approaches zero
at the crossover near $\alpha_d=0.5$ from the regime
of boosted velocities to the damped regime.

In summary, we have examined the active rheology of a single probe particle
driven though a medium with odd viscosity in the absence of
quenched disorder. We find
striking differences in the behavior
of the drag and velocity-force relations
compared to what is observed
for active rheology in overdamped systems. 
When the driving force is fixed,
we show that if the odd viscosity is sufficiently large,
the velocity of the particle
can actually increase as the density of the background particles increases,
while in the overdamped regime the opposite effect is observed
and the velocity monotonically decreases
with increasing density due to the increased frequency of
particle-particle collisions.
The velocity-force curves for
the odd viscosity system show regimes in which
the velocity is boosted to a value significantly
higher than
what is observed in the single particle limit.
This velocity boost,
which reaches a maximum value at a specific driving force,
arises due to the formation of a local density gradient
around the driven particle, which produces unbalanced repulsive forces
perpendicular to the driving direction that are converted
by the Magnus term into a velocity component parallel to the driving
direction.
At low and high drives, the size of these
density gradients and the corresponding velocity boost
are strongly reduced.
The probe particle also exhibits
a finite Hall angle which depends on density and driving force.
The Hall angle increases from zero at low drives
and approaches the single particle limit at higher drives,
similar to the behavior observed
for skyrmions driven over quenched disorder.
Our results should be general to
a wide class of
systems with odd viscosity, including skyrmions in chiral magnets. 

\acknowledgments
We gratefully acknowledge the support of the U.S. Department of
Energy through the LANL/LDRD program for this work.
This work was supported by the US Department of Energy through
the Los Alamos National Laboratory.  Los Alamos National Laboratory is
operated by Triad National Security, LLC, for the National Nuclear Security
Administration of the U. S. Department of Energy (Contract No. 892333218NCA000001).

\bibliographystyle{eplbib}
\bibliography{mybib}

\end{document}